\documentclass[pra,twocolumn,nofootinbib,floatfix]{revtex4-1}
\usepackage[utf8]{inputenc}
\usepackage[T1]{fontenc}

\usepackage{graphicx}
\usepackage{amsmath, amsxtra, amssymb, mathtools, isomath, xspace}
\usepackage{bbold}

\usepackage{hhline}
\usepackage{multirow}
\usepackage{sidecap}
\usepackage{hyperref}

\newcommand{\ket}[1]{\ensuremath{\left| #1 \right\rangle}\xspace}

\long\def\symbolfootnote[#1]#2{\begingroup%
\def\thefootnote{\fnsymbol{footnote}}\footnotetext[#1]{#2}\endgroup}

\begin{document}

\title{  
Exploring the many-body localization transition in two dimensions
}

\author{Jae-yoon~Choi$^{1\ast \dag}$}
\author{Sebastian~Hild$^{1\ast}$}%
\author{Johannes~Zeiher$^{1}$}%
\author{Peter~Schau\ss$^{1 \ddag}$}%
\author{Antonio Rubio-Abadal$^{1}$}%
\author{Tarik~Yefsah$^{1\S}$}%
\author{Vedika Khemani$^{3}$}%
\author{David~A.~Huse$^{3,4}$}%
\author{Immanuel~Bloch$^{1,2}$}%
\author{Christian~Gross$^{1}$}%

\affiliation{$^{1}$Max-Planck-Institut f\"{u}r Quantenoptik, 85748 Garching, Germany,}
\affiliation{$^{2}$Fakult\"{a}t f\"{u}r Physik, Ludwig-Maximilians-Universit\"{a}t, 80799 M\"{u}nchen, Germany}
\affiliation{$^{3}$Physics Department, Princeton University, Princeton, 08544 New Jersey, USA}
\affiliation{$^{4}$Institute for Advanced Study, Princeton, 08540 New Jersey, USA}

\symbolfootnote[1]{These authors contributed equally to this work.}
\symbolfootnote[2]{Electronic address: {\bf jae-yoon.choi@mpq.mpg.de}}
\symbolfootnote[3]{Present address: Physics Department, Princeton University, Princeton, 08540 New Jersey, USA}
\symbolfootnote[4]{Present address: Laboratoire Kastler Brossel, CNSR, Ecole Normale Sup\'{e}rieure, 75005 Paris, France}






\begin{abstract}
One fundamental assumption in statistical physics is that generic closed quantum many-body systems thermalize under their own dynamics.
Recently, the emergence of many-body localized systems has questioned this concept, challenging our understanding of the connection between statistical physics and quantum mechanics.
Here we report on the observation of a many-body localization transition between thermal and localized phases for bosons in a two-dimensional disordered optical lattice.
With our single site resolved measurements we track the relaxation dynamics of an initially prepared out-of-equilibrium density pattern and find strong evidence for a diverging length scale when approaching the localization transition.
Our experiments mark the first demonstration and in-depth characterization of many-body localization in a regime not accessible with state-of-the-art simulations on classical computers.\end{abstract}
\maketitle
Already in his seminal work on localization, Anderson emphasized the implications of localization on the thermodynamics of closed quantum systems~\cite{anderson1958}.
The absence of thermalization is truly remarkable when it persists for generic interacting systems.
Recently, new perturbative arguments suggested the existence of such non-thermalizing, many-body localized systems at low energy~\cite[and references therein]{basko2006,gornyi2005}, and soon after these arguments were extended to all interaction strengths and energy densities for systems with a bounded spectrum~\cite{aleiner2010,oganesyan2007}.
The implication, nothing less than a breakdown of equilibrium statistical mechanics for certain generic macroscopic systems, triggered tremendous theoretical efforts~\cite{nandkishore2015,altman2015a,eisert2015}.
Furthermore, the breakdown of the eigenstate thermalization hypothesis~\cite{deutsch1991,srednicki1994,tasaki1998,rigol2008} due to the failure of these systems to act as their own heat bath implies the persistence of initial state information, which might serve as a useful resource for quantum information technologies~\cite{serbyn2014}.
Several other remarkable features of many-body localization (MBL) have been uncovered, such as the description of fully localized systems by coupled localized integrals of motion~\cite{serbyn2013,huse2014}.
This underlies the absence of particle transport, but allows the ``transport'' of phase correlations, leading to a characteristic logarithmic growth of the entanglement entropy in the case of short range interactions~\cite{znidaric2008,bardarson2012,vosk2013,serbyn2013a,nanduri2014}.
Another distinctive feature as compared to non-interacting low dimensional systems is the requirement of a nonzero disorder strength for the localized phase to appear~\cite{potter2015,vosk2015}.\\
Recently, the first experimental observation of many-body localization in a quasi-disordered one-dimensional (1D) Fermi lattice has been reported~\cite{schreiber2015,bordia2016}.
These studies explored the behavior at long times and high energy density distinct to earlier experiments with non-interacting systems~\cite{wiersma1997,schwartz2007,billy2008,roati2008,kondov2011} or ultracold atoms focusing on the low energy sector~\cite{fallani2007,deissler2010,pasienski2010,gadway2011,derrico2014,tanzi2013,kondov2015,meldgin2016}.
Indications for localization in Fock space, one characteristic property of MBL~\cite{basko2006}, have been reported in short ion chains~\cite{smith2015} and MBL has been suggested as one possible explanation for the recently observed vanishing conductance in disordered superconductors at nonzero temperature~\cite{ovadia2015}. Despite intensive theoretical and experimental efforts, however, many aspects of many-body localization are still not fully understood, especially in the vicinity of the localization transition and beyond the 1D case~\cite{lev2016}.



\begin{figure*}
\centering
\includegraphics[width=1.9\columnwidth]{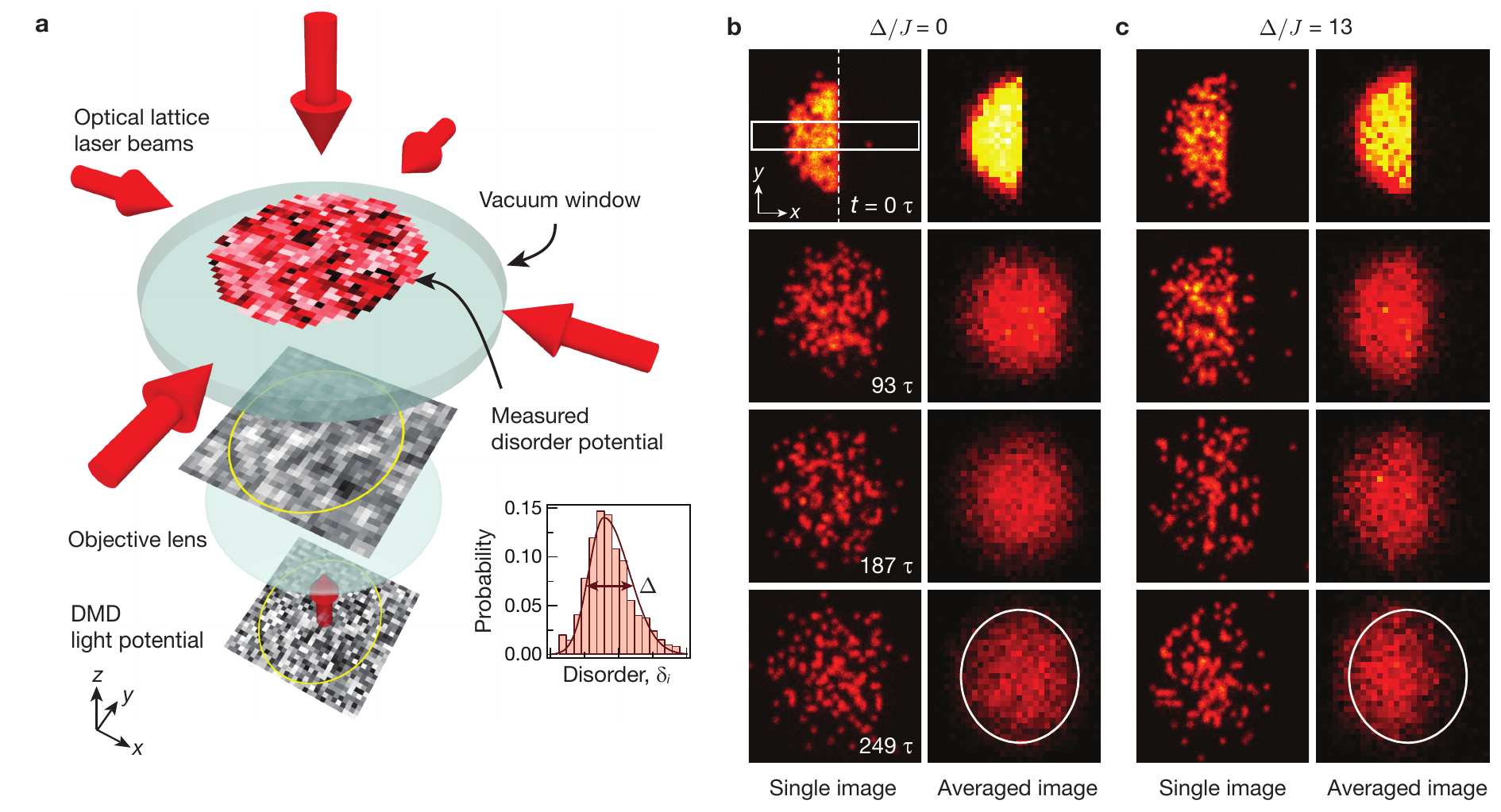}
\caption{
  \textbf{Schematics of the experiment and raw images.} \textbf{a,} A two-dimensional random disorder potential is imaged onto a single atomic plane in an optical lattice.
  The disorder is controlled by a digital mirror device (DMD), which converts a gaussian laser intensity profile into a two-dimensional random intensity distribution with spatially uniform average light intensity (bottom image).
  The limited numerical aperture (NA=$0.68$) of the microscope objective introduces a finite correlation length and leads to a smoothing of the disorder distribution.
  Lower right histogram (red bars) is the measured disorder distribution and its asymmetric gaussian fit curve (red solid line), where $\Delta$ is the full-width half-maximum of the disorder distribution.
  Distinct to the other two images showing original and smoothed (middle image) light intensity distributions, the top most image displays the local disorder potential determined by in-situ spectroscopy~\cite{som}.
  The yellow circle on the lower images indicates the spectroscopically calibrated region.
  \textbf{b,} Raw fluorescence images (red to yellow corresponds to increasing detected light level) showing the evolution of the initial density step without disorder.
  The left column shows single images (isolated red dots are individual atoms) of the parity projected atomic distribution for the indicated evolution times.
  The right column displays the mean density distribution averaged over $50$ different disorder potentials.
  The top most picture is the initial state for which the analysis region (${dx \times dy = 5 \times 31}$) is indicated by the white box.
  For the high disorder case shown in \textbf{c} the detected initial state filling is slightly lower, which is an artifact of the parity projection~\cite{som}.
  Distinct to \textbf{b}, traces of the initial state remain for all times in the disordered case. The white guide lines in the averaged density profiles after $t=249\,\tau$ highlight the difference.
}
\end{figure*}

Here, we address the open question of the existence of a many-body localization transition in two dimensions (2D), which we observe experimentally and for which we provide a first characterization.
In particular, we report on the in-situ study of thermalization and transport in a disordered 2D bosonic optical lattice. By tracking the time evolution of an initially prepared density domain wall for variable disorder strengths, we reveal the fairly sharp onset of non-thermalizing behavior above a critical value.
The observed localization transition is found when the disorder, the single particle bandwidth, the onsite interaction and the energy density are all of comparable strength.
This regime of parameters, being far from the disorder dominated (classical) and the interaction dominated limits, is highly non-trivial from the theoretical perspective.
Precise and direct characterization of the projected disorder potential by site resolved spectroscopy allows for a direct comparison of our results with a numerical prediction for the non-interacting model, highlighting the dramatic differences and pointing out the key role of interactions.
Furthermore, by locally comparing the observed density profiles to a thermalizing reference measured without disorder, we obtain strong indication for a diverging length scale when approaching the transition from the localized side.


Our experiments began by preparing a two-dimensional approximately unity filled Mott insulator of bosonic rubidium-87 atoms in a single plane of a cubic optical lattice with lattice spacing $a_{\mathrm{lat}}=532\,$nm.
We ensured the initial state to be separable by freezing out any motion at a lattice depth of $40\,E_r$ in the $x$ and $y$ directions, where $E_r=h^2/8ma_{\mathrm{lat}}^2$ is the recoil energy, $h$ the Planck constant, and $m$ the atomic mass.
Next, we optically removed the right half of the atomic population using a digital mirror device based spatial light modulator~\cite{weitenberg2011a} such that approximately $N=125(11)$ atoms remained in lattice sites located at $x<0$, thus, preparing a sharp density domain wall.
For each experimental realization a new computer-generated random disorder pattern drawn from a uniform distribution was displayed on the light modulator and subsequently projected onto the atoms, see Fig.~1a.
The optical projection results in a slightly asymmetric disorder distribution and a convolution of the disorder with the point spread function of the imaging system leads to a finite disorder correlation length $0.6~a_{\mathrm{lat}}$ and a narrowing of the distribution to a width $\Delta$~\cite{som}.
The disorder was characterized directly using single site resolved spectroscopy of the atomic sample and the topmost disorder picture in Fig.~1a displays the result of one such spectroscopic measurement for the configuration shown in the two images below.
The dynamics of the initial domain wall was then initiated by lowering the $x$ and $y$ lattice depths to $12\,E_r$ in 5~ms, which is less than one tunneling time.
Next, we allowed the system to evolve for a variable time $t$ in the disorder potential after which the local parity projected density was measured by fluorescence imaging~\cite{sherson2010}.


During the dynamics, the system is described by a two-dimensional Bose-Hubbard Hamiltonian with onsite disorder
\begin{equation}
\hat{H}=-J\sum_{\langle \mathbf{i},\mathbf{j} \rangle}\hat{a}^\dagger_\mathbf{i} \hat{a}_\mathbf{j} + \frac{U}{2}\sum_\mathbf{i} \hat{n}_\mathbf{i}(\hat{n}_\mathbf{i}-1)+ \sum_\mathbf{i} (\delta_\mathbf{i} + V_\mathbf{i}) \hat{n}_\mathbf{i}. \nonumber
 \end{equation}
Here $\hat{a}^\dagger_\mathbf{i}$ ($\hat{a}_\mathbf{j}$) is the bosonic creation (annihilation) operator, $\hat{n}_\mathbf{i}=\hat{a}^\dagger_\mathbf{i} \hat{a}_\mathbf{i}$ the local density operator on site $\mathbf{i}=(i_x, i_y)$ and the first sum includes all neighboring sites.
A harmonic trapping potential $V_\mathbf{i} = m (\omega_x^2 i_x^2 + \omega_y^2 i_y^2)/2$ with frequencies ($\omega_{x},\omega_{y})=2\pi\times (54,60)\,$Hz in $x$ and $y$ direction confines the atoms around the trap minimum.
The nearest-neighbor hopping strength at a lattice depth of $12\,E_r$ is $J/h=24.8\,$Hz corresponding to a tunneling time of $\tau=h/2\pi J=6.4\,$ms, and longer range hopping terms are neglected as they are exponentially suppressed. The onsite interaction strength is $U=24.4\,J$ and $\delta_i$ denotes the onsite disorder potential.
For these parameters, in the absence of disorder, the system's ground state is in the Mott insulating phase, however, with strong particle hole fluctuations~\cite{endres2011}.


For reference, we first tracked the evolution without any disorder potential applied.
Already from the bare images shown in Fig.~1b it becomes apparent that the initially prepared density step is smeared out after a few tens of tunneling times $\tau$ and after longer time no information about the initial density step remains.
The observed density distribution appears thermal, and neglecting quantum fluctuations at $12\,E_r$, that is, assuming decoupled sites, we extracted an upper limit of the temperature of $T<0.54(1)\,U/k_B$, where $k_B$ is the Boltzmann constant~\cite{sherson2010}.
The corresponding energy per particle of $E_{T}/N=0.58(1)\,U$ agrees with the expectation for a thermalized state.
Here, the energy density of the initial out-of-equilibrium state contributes with $E_0/ N=0.28(3)\,U$, determined by the initial thermal energy, the harmonic trap with frequency $\omega_x$ and the heating during the $2\,$s evolution time ($E_H/N=0.18(6)\,U$) used in this measurement~\cite{som}.
On the contrary, repeating the measurement with strong disorder, traces of the initial state remain and the system does not relax to a thermal state with a spatially symmetric density distribution expected for thermal state (see Fig.~1c).

\begin{figure}
\centering
\includegraphics[width=0.95\columnwidth]{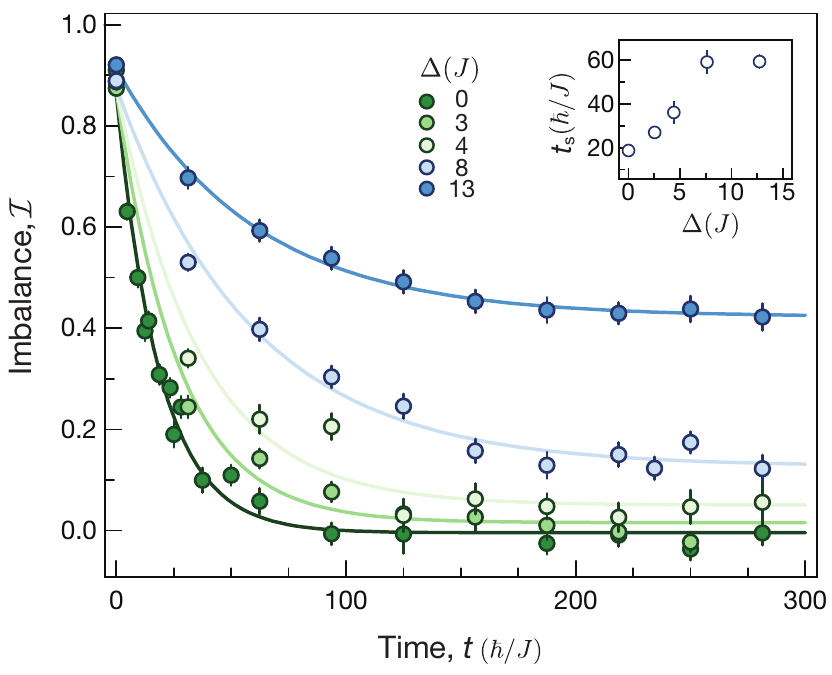}
\caption{
  \textbf{Relaxation dynamics of a density domain wall.}
  The evolution of the imbalance $\mathcal{I}$ shown for five different disorder strengths $\Delta/J=0$ (dark green), 3 (medium green), 4 (light green), 8 (light blue) and 13 (dark blue) displays a saturation behavior towards a quasi steady state for all disorder strengths.
  For low disorders (green curves) the asymptotic value of the imbalance is vanishing, while a finite imbalance remains for higher disorder (blue).
  The solid lines are fits to the data with $\mathcal{I}=\mathcal{I}_0\exp{(-t/t_s)}+\mathcal{I_{\infty}}$, of which the decay time $t_s$ is plotted versus disorder strength $\Delta$ in the inset.
  Error bars are one standard deviation of the mean in the main figure and $95\%$ confidence bounds of the fit parameters in the inset.
}
\end{figure}

A direct and model free quantity to identify a non-thermalized state is the density asymmetry quantified by a nonzero left ($N_L$) and right ($N_R$) atom number imbalance $\mathcal{I}=\frac{N_L-N_R}{N_L+N_R}$, which we analyze, as all other extracted quantities, in a central region of interest extending over $5$ lattice sites in the $y$-direction.
The zero line $x=0$, separating left and right, is defined by the position of the initial density step and was precisely aligned to the closest lattice site to the trap center resulting in an offset of up to $\pm 1$ lattice sites, corresponding to $\mathcal{I}\pm0.05$.
The evolution of the imbalance, see Fig.~2, confirms that for all disorder strengths the system reaches a quasi steady state within approximately $150\,\tau$.
For small disorder strengths we find a vanishing imbalance, while for large disorder a nonzero imbalance remains even for long evolution times.
We interpret this latter regime as the many-body localized phase, where the observed quasi steady state is clearly non-thermal and transport through the system is blocked.
The relaxation time $t_s$, extracted by an exponential fit to the data, increases with disorder strength and interestingly saturates in the non-thermal regime (see inset of Fig.~2).
We now turn to a series of measurements where we fix the evolution time to approximately $190\,\tau$, which is well in the quasi steady state regime but short enough to keep effects due to noise induced coupling to higher energy bands and atom number loss negligible.
On this timescale, we also expect the effects of low frequency noise on the disordered system to be small.
Considering the measured heating rate in the non-disordered system as an upper bound for the energy increase, the energy per particle would change by only approximately $10\%$ within one relaxation time $t_s$. Small couplings with the environment might possibly lead to relaxation of the quasi steady state on timescales much longer than our experimental time scale~\cite{nandkishore2014,fischer2015,johri2015,levi2015}.

\begin{figure}
\centering
\includegraphics[width=0.95\columnwidth]{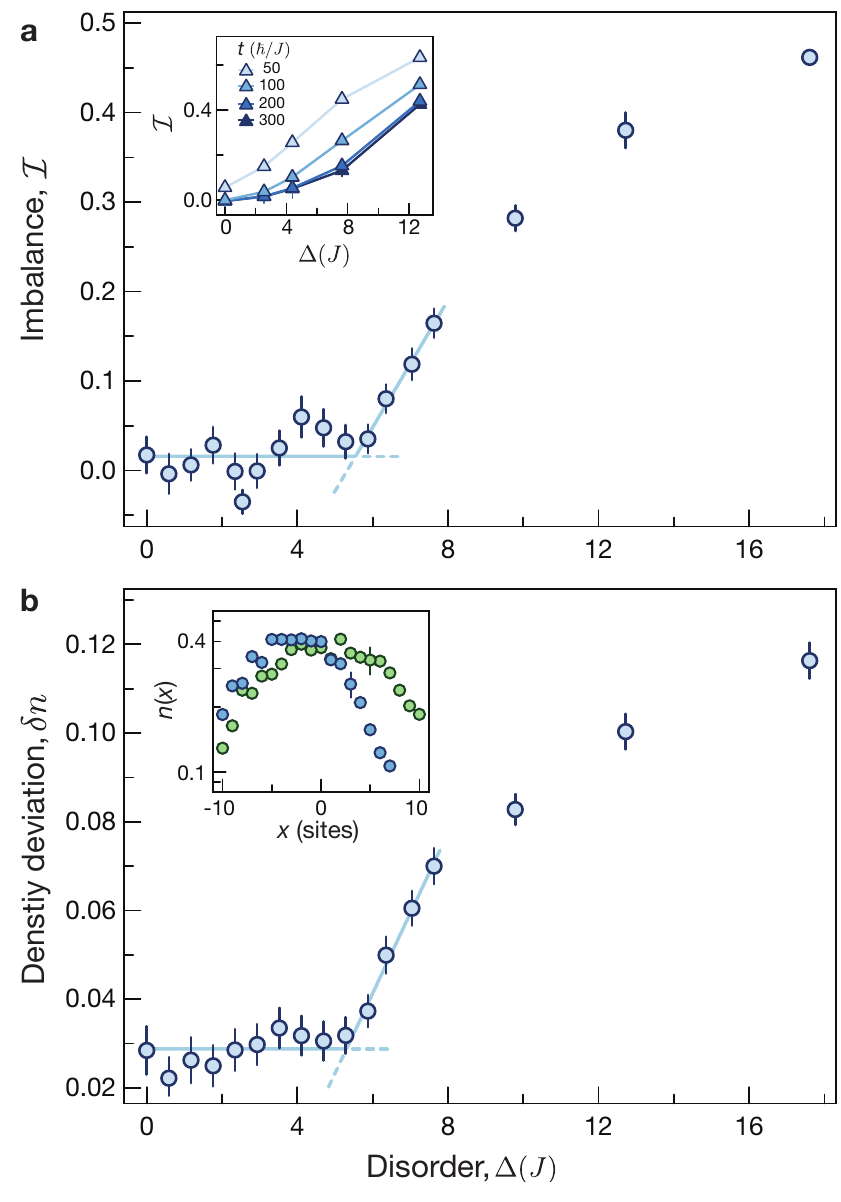}
\caption{
  \textbf{Identifying the many-body localization transition.}
  \textbf{a,} Disorder dependence of the imbalance after equilibration of the dynamics (constant evolution time of $187~\tau$, also in \textbf{b}).
  The data shows a sharp onset of nonzero quasi steady state imbalance $\mathcal{I}$ at a disorder strength of $\Delta_{c,\mathcal{I}}=5.5(4)\,J$ and the solid line is a double linear fit described in the text to extract the critical disorder.
  The inset illustrates the sharpening of the transition versus time and demonstrates the saturation of its shape by showing the imbalance extracted from the fits in Fig.~2 at different evolution times.
  \textbf{b,} Deviation from the zero disorder thermal profile measured by the root-mean-square density difference $\delta n$ at various disorder strengths.
  The relaxed density profiles differ by more than the random position induced measurement noise from the thermal profile abruptly above the critical disorder strength ${\Delta_{c,\delta n}=5.3(2)\,J}$.
  The inset shows the averaged reference density profile for zero disorder (green) and the averaged profile at a high disorder of $\Delta = 13\,J$ (blue).
  Error bars are one standard deviation of the mean.
}
\label{fig:stepdisorder}
\end{figure}

The transition from zero to nonzero imbalance $\mathcal{I}_{\infty}$ for large disorder indicates the presence of a thermalizing phase for low disorder strengths and an apparent transition to a localized phase at higher disorder.
In order to locate a many-body localization transition, we recorded a series of measurements with fixed evolution time in the quasi steady state regime and scanned the disorder around the critical value, see Fig.~3.
We find a fairly sharp onset of nonzero steady state imbalance at a disorder strength $\Delta_{c,\mathcal{I}} = 5.5(4)\,J$ indicating the transition is taking place in the system, where we extracted the critical disorder from a simple, double linear fit $\mathcal{I}(\Delta) = \mathcal{I}_0 + \mathcal{I}_1\cdot \mathrm{max}[(\Delta-\Delta_{c,\mathcal{I}}), 0]$ with $\Delta/J\in[0,8]$.
In the vicinity of the transition, slow sub-diffusive transport has been predicted~\cite{barlev2015,agarwal2015,potter2015,vosk2015,lev2016} suggesting that our measurements might not have reached a true steady state in this regime.
Because of this, it is possible that if we were able to study much longer times, the resulting transition might move towards stronger disorder.
Our high resolution detection allows for a local comparison of the measured density profiles with the equilibrated thermal profile observed at vanishing disorder.
In Fig.~3b we use this method as a more sensitive probe to detect deviations from the thermal profile by calculating the root mean square deviation $\delta n = (\sum_i [n_{i}(0) - n_{i}(\Delta)]^2)^{1/2}$ of the vertically ($y$-direction) averaged reference profile $n_{i}(0)=\frac{1}{5} \sum_{j=-{2}}^{2} n_{i,j}(0)$ and the finite disorder profiles $n_{i}(\Delta)$.
We observe that the profiles start to deviate at $\Delta_{c,\delta n}=5.3(2)\,J$ with a fairly sharp kink signaling the transition, which is quantitatively consistent with the imbalance measurements.
While in the non-interacting case, localization is predicted for vanishingly small disorder strength, the finite interaction $U$ in our system promotes thermalization. This is consistent with our observation of a thermal behavior at non-zero $\Delta$. However, as the disorder strength is increased above a critical value, the localization is restored, which is remarkable, especially from the fact that this transition takes place in the regime where the disorder $\Delta$, the interaction $U$, and the single particle bandwidth $8\,J$ are comparable.


\begin{figure}
\centering
\includegraphics[width=0.95\columnwidth]{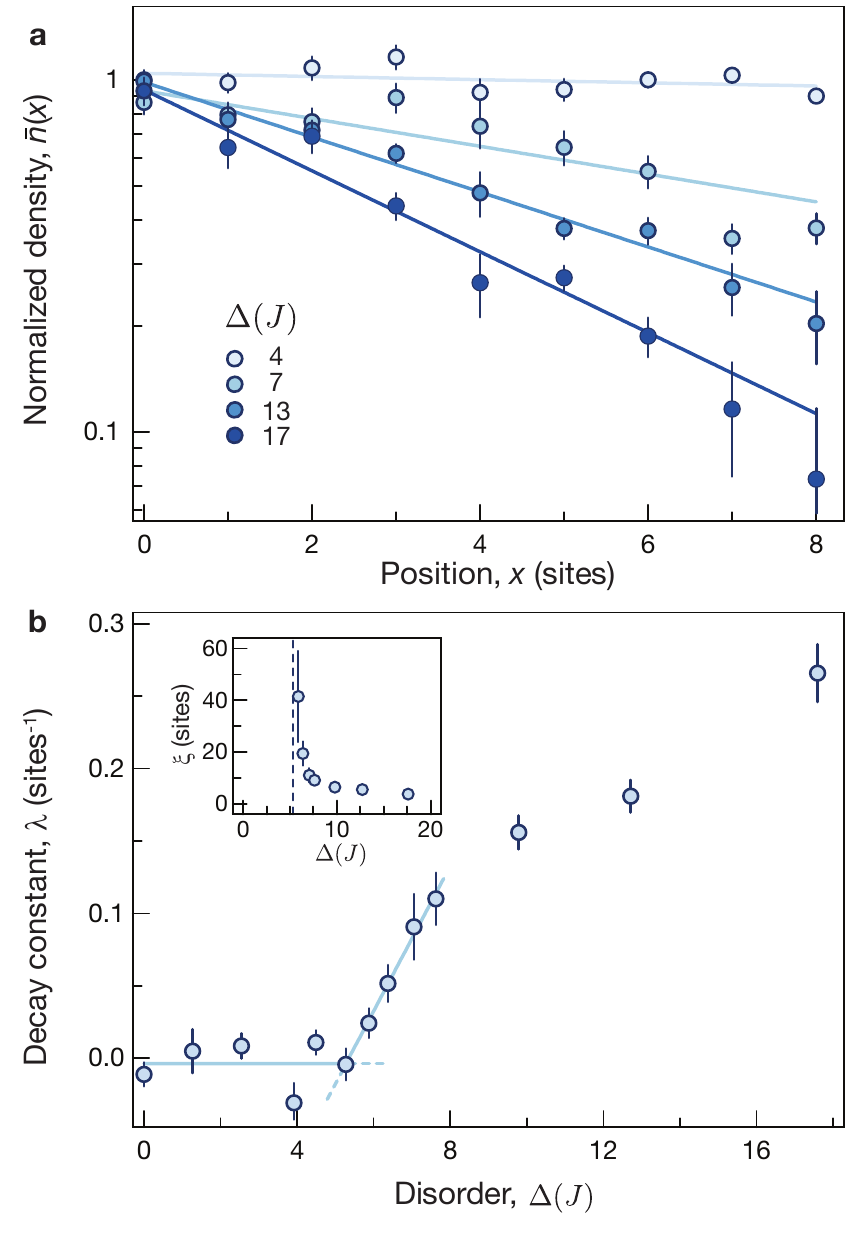}
\caption{
  \textbf{Diverging density decay length at the localization transition.}
  \textbf{a,} The spatial dependence of the normalized average density $n(\Delta)/n(0)$ in the initially empty region is fitted by an exponentially decaying model (solid lines).
  The blue brightness encodes the disorder strength increasing from light to dark: $\Delta/J=4$, $7$, $13$ and $17$.
  \textbf{b,} Fitted decay constant $\lambda$ as a function of disorder strength $\Delta$.
  The solid light blue line is a double linear fit described in the text to locate the transition point $\Delta_{c,\lambda}=5.3(2)J$.
  The inset shows the diverging decay length $\xi=1/\lambda$ near the critical disorder strength.
  Error bars are one standard deviation of the mean in \textbf{a} and $95\%$ confidence bounds of the fit parameters in \textbf{b}.
}
\end{figure}

The MBL phase transition is expected to be a continuous transition~\cite{pal2010,kjall2014,vosk2015,potter2015,chandran2015} for which one expects a characteristic diverging length scale when approaching the transition.
With our experiments, we have direct access to the steady state decay length $\xi(\Delta)$ of the initially prepared density step, which is directly related to a density-density correlation length.
To minimize the influence of the external trap, we reference the density profile $n_i(\Delta)$ to the thermal profile $n_i(0)$ by calculating $\bar{n}_i(\Delta)=n_i(\Delta)/n_i(0)$, see Fig.~4a.
The observed decay is found to be well captured by an exponential fit $\bar{n}_i(\Delta)\sim e^{-\lambda(\Delta) i}$ with a disorder dependent decay constant $\lambda(\Delta)=1/\xi(\Delta)$.
A priori, we would have expected a crossover from a interaction dominated decay to a single particle dominated decay at low densities present at the outer edge of our sample, however, within the experimental uncertainty we cannot identify such an effect.
We directly observe the diverging behavior of the decay length $\xi(\Delta)$ when approaching the transition from the localized side, see inset of Fig.~4b.
The related decay constant $\lambda(\Delta)$, also shows a very sharp kink at $\Delta_{c,\lambda}=5.3(3)\,J$, marking the onset of the localized region.
We empirically obtain the critical disorder $\Delta_{c,\lambda}$ using the bilinear fit function, $\lambda(\Delta)=\lambda_0+\lambda_1\cdot\max[(\Delta-\Delta_{c,\lambda}),0]$ with $\Delta/J\in[0,8]$.
We emphasize, that all three methods to extract the critical disorder strength of the MBL transition in the experiments agree within the fit errors.

\begin{figure}
\centering
\includegraphics[width=0.95\columnwidth]{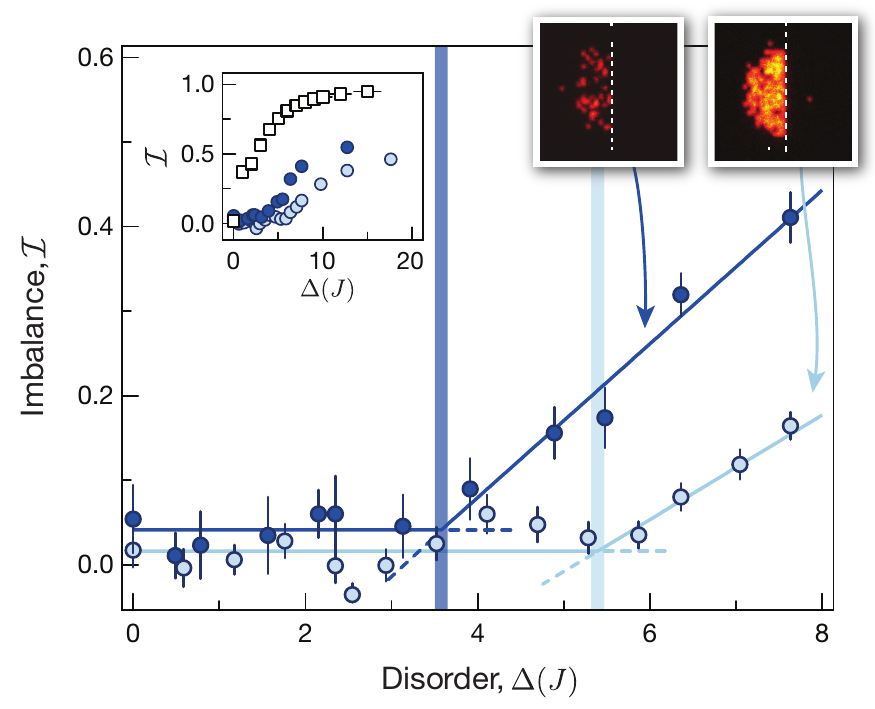}
\caption{
  \textbf{Interaction dependence of the localization transition}
  Quasi steady state imbalance $\mathcal{I}$ versus disorder strength $\Delta$ for different initial densities.
  The interaction effects are reduced by lowering the initial filling to $0.23$, which is $25\%$ of the value previously discussed in Fig.~3a (light blue).
  The clear difference in critical disorder strengths highlights the strong influence of interactions on the localization.
  The right two insets show representative fluorescence images of the initial density distribution for each case.
  The left inset is a zoomed out view of the main figure where we added the results of exact diagonalization numerics for the non-interacting system with the same experimental conditions (black open squares).
  Here, the horizontal error bars denote the systematic uncertainty in the disorder strength.
  Vertical error bars are one standard deviation of the mean.
}
\end{figure}

In order to experimentally verify that the observed behavior of the system is induced by interactions, we reduced the initial density and therefore the effects of the interaction on the dynamics while keeping the initial energy per particle $E_0/N$ constant.
Lower atomic densities are obtained by uniformly transferring a given fraction of the atoms by applying a microwave pulse to another hyperfine state and subsequently removing optically the transferred population.
When reducing the density by factor of four, we observe a clear shift of the localization transition towards lower disorder strength as is expected, see Fig.~5.
Here, the left-right imbalance $\mathcal{I}(\Delta)$ displays a sharp transition behavior at a smaller critical disorder of $\Delta_{c,\mathcal{I}}\simeq3.6(2)\,J$.
Furthermore, the numerical prediction for a non-interacting system obtained by exact diagonalization~\cite{som} is fully incompatible and strongly differs from both measurements, showing that the interactions facilitate thermalization for low disorder strengths.
Reducing the system size by preparation of a smaller initial Mott insulator did not affect the observed critical disorder~\cite{som}.
From these observations, we conclude that the nonzero critical disorder strength is due to the interactions and that the measured critical disorder value is not strongly influenced by the small external driving due to laser fluctuations discussed above.


In conclusion, our experiments provide the first evidence for many-body localization in two dimensions by the observation of a transition from a thermalizing phase to a localized phase of interacting bosons in a disordered optical lattice.
The system size analyzed in our experiment is far beyond numerically accessible scales, demonstrating a non-trivial quantum realization of the MBL system that challenges both analytical and numerical theory.
Furthermore, we supplemented our observation of a MBL transition with the demonstration of a clear shift of the transition point for effectively smaller interaction energy.
Even though it is difficult to distinguish a true phase transition from a sharp crossover within experiments, our results mark a first step in understanding many-body localization in more than one dimension, and can be extended to obtain detailed information about the nature of the MBL transition, such as its dynamical critical exponent~\cite{potter2015,vosk2015}.
Furthermore, supplementing transport experiments with density-density correlation measurements offer a promising possibility to demonstrate ongoing phase dynamics in the localized phase while the density or charge transport is frozen~\cite{smith2015,goihl2016}.
By detailed studies of the dynamics of transport it should also be possible to study Griffiths effects and sub-diffusive transport in the vicinity of the transition~\cite{barlev2015,agarwal2015,potter2015,vosk2015,lev2016}.
Besides dynamical properties our technique might allow to probe many-body eigenstate related properties, such as the local integrals of motion~\cite{serbyn2013,huse2014} defined via local operators~\cite{chandran2015}.
\section*{Acknowledgements}
We thank M. Fischer, E. Altman, M. Knap, P. Bordia, H. L\"{u}schen, A. Rosch, E. Demler, and S. Sondhi for discussions. D.A.H. is the A. and H. Broitman Member at I.A.S. We acknowledge funding by MPG, DFG, EU (UQUAM, Marie Curie Fellowship to J.C.).




\bibliography{referencesArXiv}



\clearpage
\newpage
\section*{Supporting Material}

\section{Initial state preparation and detection of the atoms}

To prepare the density domain wall, we started with a two-dimensional Mott insulator of $0.90(5)$ parity projected central density deep in the atomic limit at $40\,E_r$ in $x$ and $y$ direction.
Next, we used $\sigma^-$ polarized laser light with a wavelength of $787.55\,$nm to induce a differential light shift of $h\times10\,$kHz between the $\ket{F, m_F} = \ket{1,-1}$ and $\ket{2,-2}$ states for the atoms in the left half of the system~\cite{weitenberg2011a}.
A microwave sweep selectively transferred the non-illuminated atoms in the right half to the $\ket{2,-2}$ state before they were removed using resonant light on the cycling transition of the D2 line.
Finally, we transferred all remaining atoms to $\ket{2,-2}$ before adiabatically turning on the $\sigma^-$ polarized disorder potential at $787.55\,$nm.
The whole preparation took $90\,$ms after which we initiated the domain wall dynamics by ramping down the lattices to $12\,E_r$ in $5\,$ms.

In order to detect the atoms after the dynamics, the lattice depth was suddenly increased to $40\,E_r$, freezing out all motion and, subsequently, the parity projected density distribution was measured by single-site resolved fluorescence imaging~\cite{sherson2010}.
Information about multiply occupied sites is lost due to the parity projection, however, the effect is small in the presented experiments.
We expect occupation numbers of three or more to be negligible and from the decrease of the detected atom number after the evolution, we estimate the fraction of doubly occupied sites to be $~2\%$ in the uniform system and $~9\%$ for strong disorder of $\Delta=17\,J$.

\section{Disorder potential}


\begin{figure}
\includegraphics[width=0.9\columnwidth]{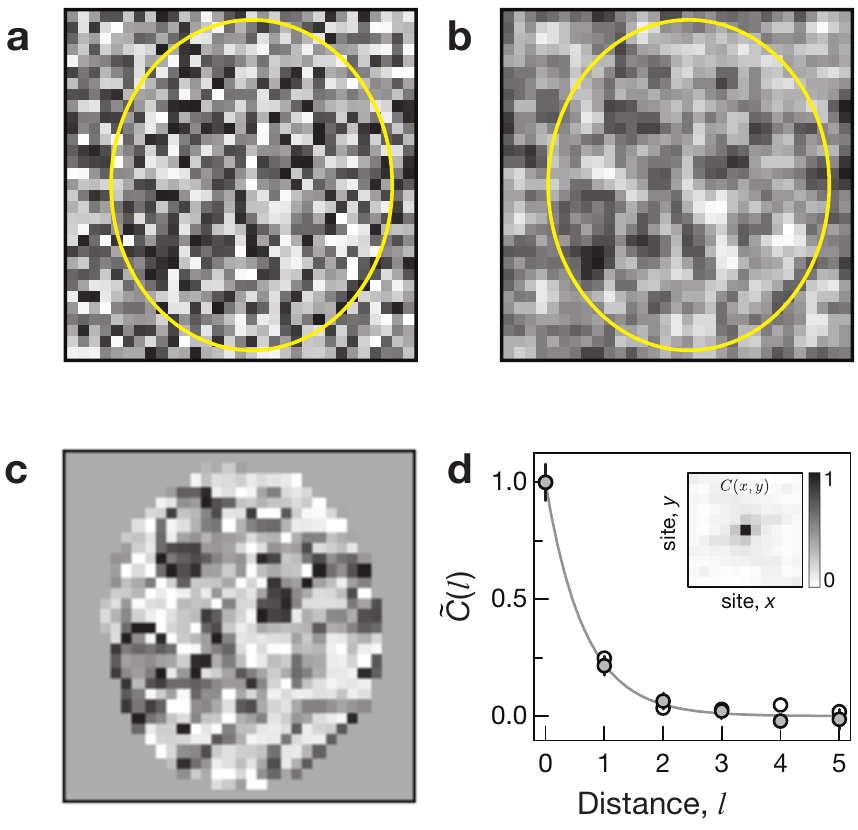}
\caption{
  \textbf{Characterization of the two-dimensional disorder.} \textbf{a,} Image of the two-dimensional random pattern ($31\times31$ lattice sites) displayed on the digital micro-mirror device.
\textbf{b,} The same random picture after convolution with the point spread function of our imaging system.
\textbf{c,} Spectroscopically measured disorder potential for the picture shown in \textbf{a}. Every point corresponds to single lattice site.
\textbf{d,} Averaged $x$ and $y$ cuts $\tilde{C}$ of the measured disorder correlation (black filled circles).
The corresponding full autocorrelation $C(x,y)$ is shown in the inset.
The solid line is an exponential fit to the profile, see text.
Open circles correspond to the expected profile taking our finite resolution into account.
Error bars represent one standard deviation of the mean.
}
\label{fig:DecayVsDisorderSI}
\end{figure}

The two-dimensional random potential was generated using a digital micro-mirror device (DMD) with $1024\times768$ binary pixels and a micro-mirror  size of $13.7\,\mu$m.
Approximately $7\times7$ of these mirrors were focused to one lattice site, which gave us the ability to create grayscales.
We used pseudo-random numbers to generate a two-dimensional random pattern, which is then transferred to the DMD (Fig.~6a).
For the creation of the disordered light pattern we reflected a gaussian shaped laser beam from the DMD surface, which was projected onto the atoms through the high-resolution objective with a numerical aperture of $0.68$.
We multiplied a global profile to the disorder pattern to cancel the intensity distribution of the initial gaussian laser beam.
The DMD surface acts as an optical grating and the following optics as a spatial low pass filter.
Depending on the displayed micro-mirror pattern, which depends on the set gray value per lattice site, the Fourier composition of the reflected light changes, which led to the observed asymmetry in the disorder distribution~\cite{dorrer2007}.
Furthermore, the low-pass filtering smoothed the disorder distribution and induced a finite correlation length of $\zeta$ in the disorder potential seen by the atoms.

We precisely characterized the disorder potential by spectroscopically measuring the light shift at each lattice site via microwave spectroscopy between the $\ket{1,-1}$ and $\ket{2,-2}$ states.
The measured local resonance shift at each lattice site $\nu(x,y)$ thus directly corresponds to the local random potential (Fig.~6c).
From this measurement, we analyzed the auto-correlation function $C(x,y)$ and extracted the finite correlation length of the random potential.
Here, $C(x,y)$ is calculated from
\begin{equation}
C(x,y)=\left<\left<\Delta \nu(x+x',y+y')\Delta \nu(x',y')\right>_{x',y'}\right>,
\end{equation}
where $\Delta \nu(x,y)=\nu(x,y)-\nu_m$ is the fluctuation of the light potential at each lattice site with respect to the mean potential depth $\nu_m$ and the outer bracket indicates averaging over four different disorder patterns.
Averaging $C(x,y)$ over the vertical or horizontal direction reveals the auto-correlation profiles $\widetilde{C}$.
These profiles are found to decay exponentially $\widetilde{C}=\widetilde{C}_0\exp{(-l/\zeta)}$ with distance $l$ and a correlation length $\zeta=0.6(0.1)~a_{\mathrm{lat}}$.
The measured correlation length agrees well with the simulated value $\zeta_{NA}=0.64~a_{\mathrm{lat}}$ due to the finite resolution of the imaging system.

For the measurements reported here it is crucial to ensure a homogeneous average disorder light level across the cloud and in particular to rule out the presence of large scale intensity gradients perpendicular to the density domain wall.
We confirmed that the overall uniformity of our light pattern is better than $5\%$ by shining a flat intensity distribution onto the atoms and measuring the local potential.
Furthermore, we tested the uniformity by evolution of the many-body system in a strong ($V_0=50\,J$) uniform offset potential and found a symmetric steady state distribution in the trap.
Moreover, we compared the steady state density profiles after evolution in a disorder potential for the cases of an initial state prepared on the left and on the right and found no significant difference.

\section{Heating dynamics of Mott insulator}

\begin{figure}
\centering
\includegraphics[width=0.95\columnwidth]{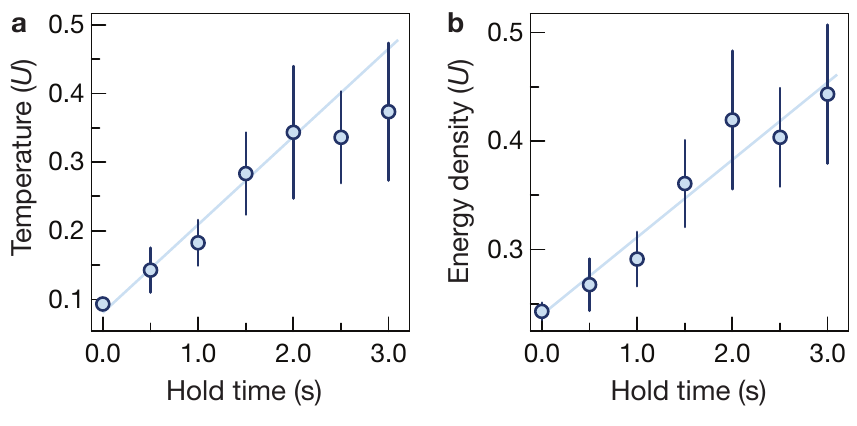}
\caption{
\textbf{Heating of a Mott insulator.} \textbf{a,} Temperature and \textbf{b,} corresponding energy density of an initially approximately unity filled Mott insulator for various hold times. The solid lines are linear fits to the data and the error bars are one standard deviation of the mean.
}
\label{fig:HeatingS2}
\end{figure}

Fluctuations of the optical lattice pointing and intensity result in heating of the many-body system which is dependent on the spectral characteristics of the noise and the many-body system.
To estimate the energy increase during the evolution time we measured the heating rate at $12\,E_r$ lattice depth assuming decoupling of the lattice sites (atomic limit)~\cite{sherson2010}.
The decoupling assumption might lead to a small overestimation of the temperature due to quantum mechanical particle-hole fluctuations present in the system.
As shown in Fig.~7, the temperature and energy density increased approximately linearly upto 2~s.
The heating rate was obtained by a linear fit, which yielded $\dot{T}=0.11(2)\,U/s$ and a corresponding rate for the energy density $\dot{E}/N=0.09(3)\,U/s$.
Here, $U=h\times601\,$Hz is the onsite interaction energy for $12\,E_r$ lattice depth.
This heating rate estimate was measured close to the superfluid -- Mott insulator transition point, where the system is most susceptible to external noise.
Thus, we expect it to be an upper bound for the coupling to external noise sources in the case of non-vanishing disorder.

\section{System size effect}

\begin{figure}
\centering
\includegraphics[width=0.95\columnwidth]{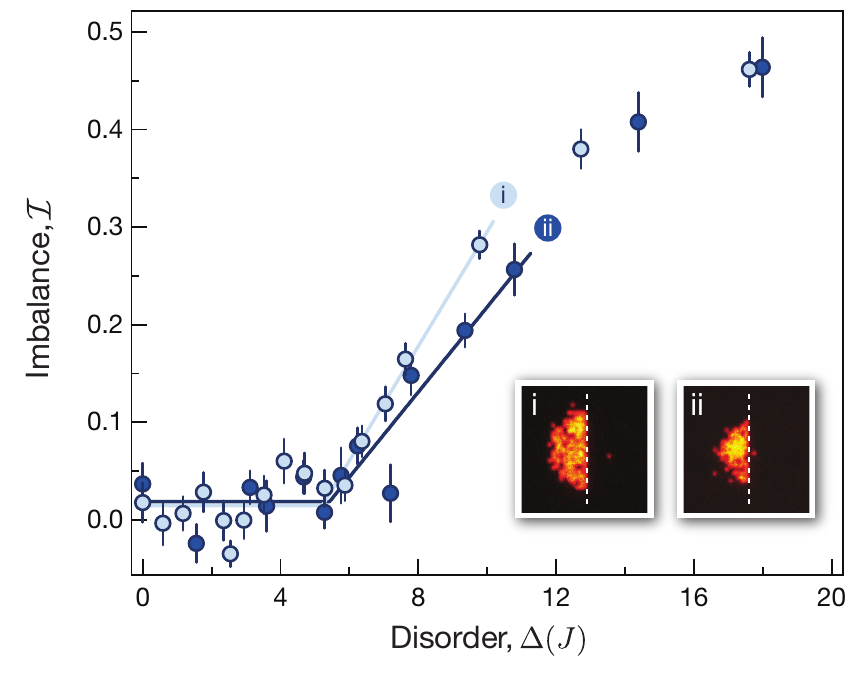}
\caption{
\textbf{Localization transition for two system sizes.}
 Imbalance $\mathcal{I}$ as function of disorder $\Delta$ for two different system sizes.
 We changed the initial radius of the Mott insulator from $9$ (i, light blue) to $7$ (ii, dark blue) lattice sites by tightening the external trap,
 but observed no change in the critical disorder strength.
 The insets show a representative fluorescence image of the initial density distribution for each case.
 Error bars are one standard deviation of the mean.
}
\label{fig:SystemsizeS3}
\end{figure}

To check for finite size effects we studied the localization transition in a smaller system.
We reduced the initial atom number by $60\%$ to $N=75(10)$ and the radius of the initial Mott insulator shrunk from $9$ to $7$ sites.
To keep the energy per particle $E_0/N$ constant, we increased the trap frequency to $\omega_x=2\pi\times65\,$Hz.
As shown in Fig.~8, there is no visible change in the critical disorder extracted from the imbalance measurements.
The fitted critical disorder for the small system is $\Delta_{c,\mathcal{I}}=5.4(6)$,
consistent with the value $\Delta_{c,\mathcal{I}} = 5.5(4)\,J$ found for the larger system.

\section{Numerics for non-interacting atoms}

\begin{figure}
\centering
\includegraphics[width=0.9\columnwidth]{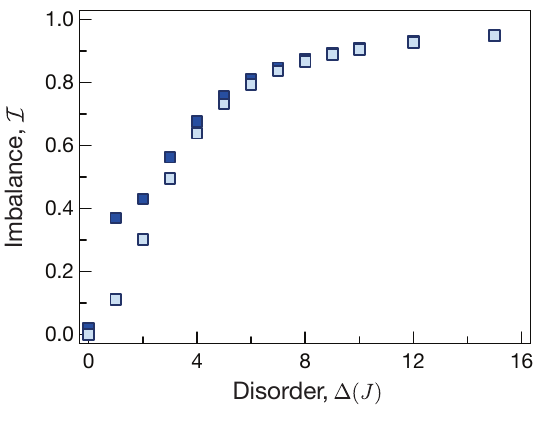}
\caption{
\textbf{Numerics for non-interacting atoms.}
Imbalance $\mathcal{I}$ as function of disorder strength $\Delta$ simulated with the confining harmonic potential (dark blue) compared to a system without trap (light blue).
The imbalances for both configuration agree well with each other at strong disorder. The imbalance shows a slight difference at lower disorder because of the harmonic trap, which induces a level shift of the lattice potential and can localize particles more easily under weak disorder.
All simulations were averaged over at least $100$ disorder realizations and the standard deviation of the mean is smaller than the point size.
}
\label{fig:SimulationS4}
\end{figure}

We used exact diagonalization to simulate the behavior of non-interacting atoms in the disordered optical lattice.
The simulations optionally include the harmonic confinement with a trapping frequency of $\omega_x=54\,$Hz as present in the experiment.
We chose a square lattice with $31\times31$ lattice sites, the same as the size of the disorder pattern in the experiment.
The experimentally realized disordered intensity distributions were modeled by squaring a two dimensional array of random numbers between zero and one and afterward convolving it with a normalized two dimensional Gaussian with a standard deviation of $0.5\,a_\mathrm{lat}$.
This reproduces the measured correlations and the distribution of the onsite energies of the experimental disorder realizations (Fig.\,1).
Finally, the disorder strength was set by multiplying the obtained distribution by a global scaling factor.
All simulations were repeated for $100$ different disorder realizations and the observables were averaged over these realizations.
We verified that the observables depend only slightly on the exact shape of the disorder distribution when fixing the spatial correlations of the patterns.
In particular, we compared the experimentally measured distribution to a symmetric gaussian distribution with the same full width at half maximum and found little difference.
Our simulations started from an initial state with all lattice sites in the left half of the system occupied that have a geometric distance below $9$ lattice sites from the center.
This corresponds to $133$ atoms and agrees well with the experimental studied clouds.
The imbalance was extracted from the central $5$ lines perpendicular to the domain wall,
exactly in the same way as it was done for the experimental measurements.
The confining harmonic dipole potential only influences the imbalance at lower disorder strength, where the trap induced onsite energy difference dominates over the disorder energy shifts (Fig.~9).


\end{document}